\documentclass[twocolumn,aps,nofootinbib]{revtex4}
\usepackage{amsmath} \usepackage{graphicx} \usepackage{dcolumn}
\usepackage{bm} \usepackage{epsf} \usepackage{amssymb}

\begin{document}

\newlength{\figurewidth}
\setlength{\figurewidth}{0.95 \columnwidth}

\newcommand{\prtl}{\partial}
\newcommand{\la}{\left\langle}
\newcommand{\ra}{\right\rangle}
\newcommand{\dla}{\la \! \! \! \la}
\newcommand{\dra}{\ra \! \! \! \ra}
\newcommand{\we}{\widetilde}
\newcommand{\smfp}{{\mbox{\scriptsize mfp}}}
\newcommand{\smp}{{\mbox{\scriptsize mp}}}
\newcommand{\sph}{{\mbox{\scriptsize ph}}}
\newcommand{\sinhom}{{\mbox{\scriptsize inhom}}}
\newcommand{\sneigh}{{\mbox{\scriptsize neigh}}}
\newcommand{\srlxn}{{\mbox{\scriptsize rlxn}}}
\newcommand{\svibr}{{\mbox{\scriptsize vibr}}}
\newcommand{\smicro}{{\mbox{\scriptsize micro}}}
\newcommand{\scoll}{{\mbox{\scriptsize coll}}}
\newcommand{\sth}{{\mbox{\scriptsize th}}}
\newcommand{\seq}{{\mbox{\scriptsize eq}}}
\newcommand{\teq}{{\mbox{\tiny eq}}}
\newcommand{\sinn}{{\mbox{\scriptsize in}}}
\newcommand{\suni}{{\mbox{\scriptsize uni}}}
\newcommand{\tin}{{\mbox{\tiny in}}}
\newcommand{\scr}{{\mbox{\scriptsize cr}}}
\newcommand{\tstring}{{\mbox{\tiny string}}}
\newcommand{\sperc}{{\mbox{\scriptsize perc}}}
\newcommand{\tperc}{{\mbox{\tiny perc}}}
\newcommand{\sstring}{{\mbox{\scriptsize string}}}
\newcommand{\stheor}{{\mbox{\scriptsize theor}}}
\newcommand{\sGS}{{\mbox{\scriptsize GS}}}
\newcommand{\sBP}{{\mbox{\scriptsize BP}}}
\newcommand{\sNMT}{{\mbox{\scriptsize NMT}}}
\newcommand{\sbulk}{{\mbox{\scriptsize bulk}}}
\newcommand{\tbulk}{{\mbox{\tiny bulk}}}
\newcommand{\cN}{{\cal N}}
\newcommand{\cB}{{\cal B}}
\def\Xint#1{\mathchoice
   {\XXint\displaystyle\textstyle{#1}}%
   {\XXint\textstyle\scriptstyle{#1}}%
   {\XXint\scriptstyle\scriptscriptstyle{#1}}%
   {\XXint\scriptscriptstyle\scriptscriptstyle{#1}}%
   \!\int}
\def\XXint#1#2#3{{\setbox0=\hbox{$#1{#2#3}{\int}$}
     \vcenter{\hbox{$#2#3$}}\kern-.5\wd0}}
\def\ddashint{\Xint=}
\def\dashint{\Xint-}
\title{Theory of Structural Glasses and Supercooled Liquids}

\author{Vassiliy Lubchenko} \affiliation{Department of Chemistry,
  University of Houston, Houston, TX 77204-5003} \author{Peter
  G. Wolynes} \affiliation{ Department of Chemistry and Biochemistry
  and Department of Physics, University of California at San Diego, La
  Jolla, CA 92093-0371}

\date{\today}

\begin{abstract}

  We review the Random First Order Transition Theory of the glass
  transition, emphasizing the experimental tests of the theory. Many
  distinct phenomena are quantitatively predicted or explained by the
  theory, both above and below the glass transition temperature
  $T_g$. These include: the viscosity catastrophe and heat capacity
  jump at $T_g$, and their connection; the non-exponentiality of
  relaxations and their correlation with the fragility; dynamic
  heterogeneity in supercooled liquids owing to the mosaic structure;
  deviations from the Vogel-Fulcher law, connected with strings or
  fractral cooperative rearrangements; deviations from the
  Stokes-Einstein relation close to $T_g$; aging, and its correlation
  with fragility; the excess density of states at cryogenic
  temperatures due to two level tunneling systems and the Boson Peak.

\end{abstract}

\maketitle

\tableofcontents

\section{Introduction}

School children have their earliest exposure to the subject of
physical chemistry when they hear about ``the states of matter.'' They
are taught there are gases, liquids, and solids. Van der Waals
revealed to scientists that gases and liquids differed actually only
quantitatively \cite{VdW}. The rigidity of solids, which defines them
macroscopically, on the other hand, has been usually traced to their
qualitatively different, periodic ordered structure
\cite{PWA_Concepts} - an idea that already occurred to Kepler
\cite{Kepler}. This idea is correct for crystals. The existence of
structural glasses, i.e. amorphous substances that are rigid, calls
this understanding into question, however. The utility and
adaptability of glasses arises from the way their properties depend on
their preparation history and their seeming continuity with the
supercooled liquid state. In contrast to dilute gases and crystalline
solids, where the properties can be directly inferred from the
intermolecular forces, liquids have generally enjoyed a reputation of
mystery among physical chemists. Until recently, the transition to the
``glassy state'' was deemed specifically one of the most obscure
enigmas by many in the theoretical physics and chemistry communities
\cite{PWA_lightly, tenQ, Angell_PNAS}. Despite this reputation, a
constructive molecular theory of structural glasses and the dynamics
of supercooled liquids {\em has} been developed. This theory starts
with the intermolecular forces. While it would suffer from all the
well-known issues of microscopic modeling, if it were used to predict
glass transition temperatures from scratch, for example, the theory
does explain extremely well the known (and rather unusual)
phenomenology of supercooled liquids and the strange properties of
glasses at low temperatures. The theory also does make a large number
of quantitative predictions without adjustable parameters that are
borne out quite well by experiment. The purpose of this article is to
briefly explain the basic ideas of this theory and how its predictions
compare with experiment.

The basic idea of the theory of structural glasses is to consider them
to be aperiodic crystals. Since Bernal's time at least, it has been
appreciated that an aperiodic structure can be mechanically stable
\cite{Bernal}. Strictly speaking, Kepler's assumption of periodicity
would not be needed for rigidity, were it not for thermal motions. In
contrast to periodic crystals, there are extremely many aperiodic
structures and it is hard to see why they should have any difficulty
interconverting thereby allowing flow. But we know they indeed do have
such a difficulty since the flow of glass is nearly imperceptible. It
is this fact that a theory of the glass transition must explain. We
shall see the theory of structural glasses then generalizes the theory
of an ordinary first order freezing transition, which gives a periodic
crystalline structure, to a very much analogous change that involves
freezing into a set of aperiodic structures which are statistically
distributed in energy. We call this change a ``random first order
transition.''

\begin{figure}[t]
\includegraphics[width=.75\figurewidth]{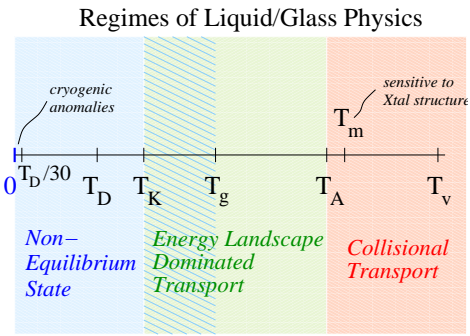}
\caption{\label{chart} Regimes of the aperiodic condensed molecular
  phase are shown, ranging between a dilute gas and a frozen glass.
  $T_v$ is the vaporization temperature, $T_m$ the melting point.
  $T_A$ represents the temperature signalling the crossover to
  activated motions, which is usually but not always below
  $T_m$. $T_g$ is the glass transition temperature which depends on
  the time scale of measurement. Below $T_g$ the system is out of
  equilibrium and ages. $T_K$ is the Kauzmann temperature (see
  text). $T_D$ is the Debye temperature which signals the quantization
  of vibrational motions. Below $T_D/30$, or so, the thermal
  properties of the system can be phenomenologically described as
  arising from a collection of two level systems. Just above this
  point, additional quantum excitations, sometimes called the Boson
  peak, are present.}
\end{figure}

We will begin this review by discussing a small number of key
experimental signatures of the glass transition in Section II. In
Section III, we construct the microscopic picture of the glassy state
and the transition to it from a supercooled liquid, following the
random first order transition theory. A variety of temperatures
characterizes glasses and liquids in this theory. They are graphically
summarized in Fig.\ref{chart}. We will define these scales more
precisely in the discussion below and we recommend the reader to often
refer to this figure. Starting with a one-component gas, one may cool
it down and compress it until it condenses below the critical point,
$T_v$, usually above the crystallization temperature $T_m$. In this
temperature range, an effective description in terms of collisional
transport is valid: a liquid is just a very dense gas held together by
an average attractive force. No two molecules are likely to reside
near each other for any significant time. The time scales for
molecular permutations and collisions are comparable in this regime.
All the pertinent information about particle-particle interactions may
be encoded in low order correlation functions that may be computed or
extracted experimentally from scattering experiments. In a supercooled
liquid, on the other hand, molecules maintain their immediate set of
neighbors for hundreds of collisional or vibrational periods. This
occurs near the temperature $T_A$. These local spatial patterns
persist ever longer as the temperature is lowered. Interconversion
between such structures occurs both above and below the glass
transition temperature $T_g$, which depends on the preparation time
scale. The interconversion is called the $\alpha$-relaxation when the
material remains in equilibrium. However, when $\alpha$-relaxation
becomes too slow and only a fraction of the interconversions have time
to occur, the material is a glass that ``ages''. Even at cryogenic
temperatures (liquid He and below), a certain fraction of the sample
will harbor several kinetically accessible states. Interconversions
can still occur by tunneling. These quantum motions are discussed in
Section IV. In the final Section V, we make concluding remarks and
highlight some open questions in the field.

\section{Basic Phenomenology of the Structural Glass Transition}
\label{Phenomenology}

\begin{figure}[t]
  \includegraphics[width=.95\figurewidth]{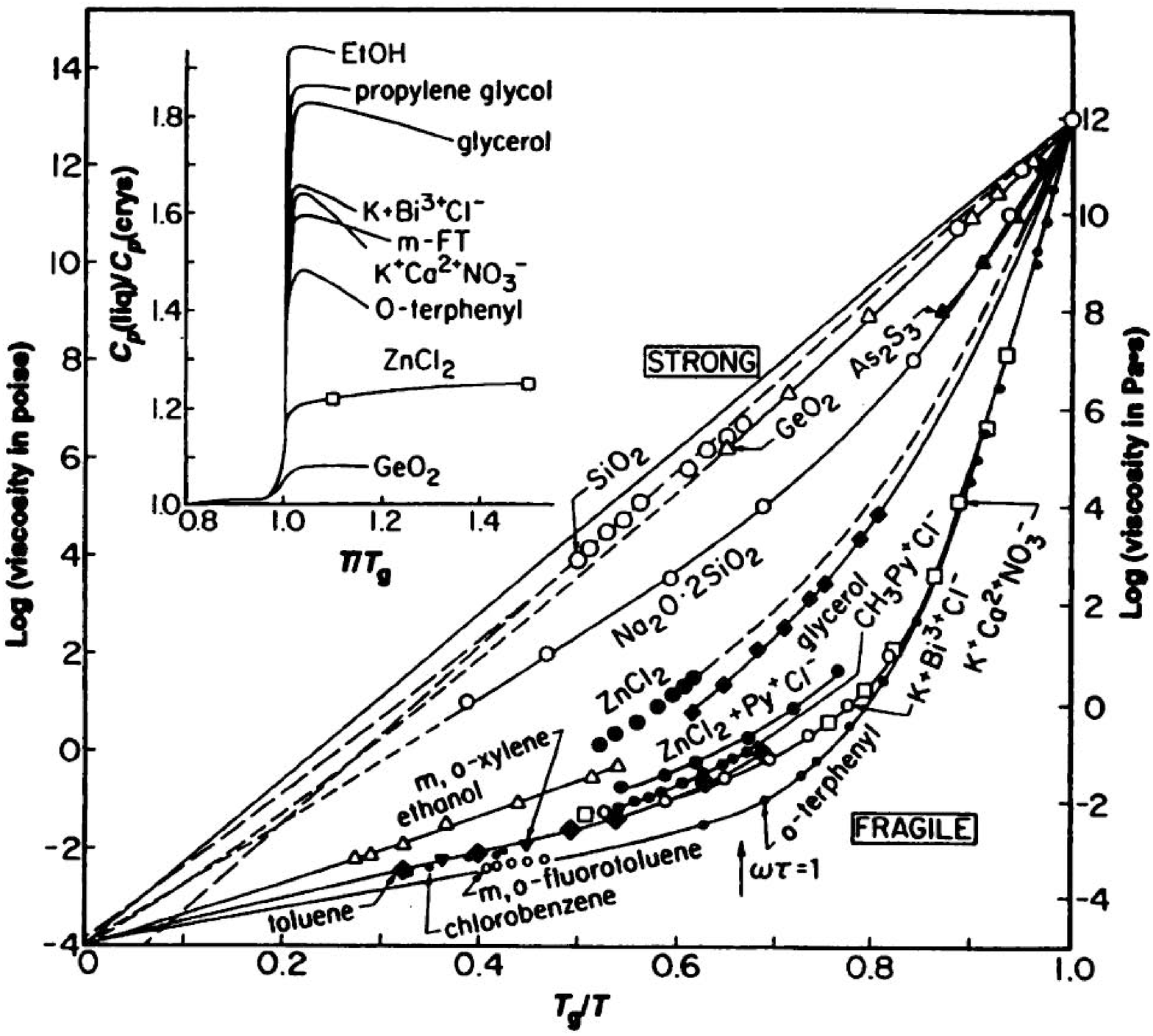}
  \caption{\label{angell} The viscosities of several supercooled
    liquids are plotted as functions of the inverse
    temperature. Substances with almost-Arrhenius-like dependences are
    said to be strong liquids, while the visibly convex curves are
    described as ``fragile'' substances. The full dynamic range from
    about a picosecond, on the lower viscosity side, to $10^4$ seconds
    or so when the viscosity reaches to $10^{13}$ poise. This figure
    is taken from Ref.\cite{Angell_PNAS}.}
\end{figure} 

Liquids exhibit a remarkable range of dynamical behaviors within a
relatively narrow temperature interval.  Viscosity, for example,
varies over a tremendous dynamic range: Fig.\ref{angell} reproduces
the celebrated ``Angell'' plot of the viscosities for superooled
liquids as functions of the inverse temperature scaled to their
respective glass transition temperatures, where the relaxation time is
roughly one hour \cite{Angell_PNAS}. The temperature dependence of
other structural relaxation times, such as the inverse of the lowest
frequency peak of the dielectric susceptibility, follow a similar
temperature dependence and can be described by the so-called
Vogel-Fulcher (VF) law, to a first approximation:
\begin{equation} \label{VF}
  \tau = \tau_0 e^{D T_0/(T - T_0)},
\end{equation}
where the material coefficient $D$ is called the liquid's
``fragility''.  The Vogel-Fulcher fits work better in the vicinity of
the temperature $T_0$, which is also a material dependent quantity.
Near $T_0$, the relaxation times grow most rapidly and would appear to
diverge, if one had the patience to equilibrate the liquid for
cosmological times. Deviations from the VF law occur, of course, when
the substance falls out of equilibrium near the laboratory $T_g$. 

\begin{figure}[t]
  \includegraphics[width=\figurewidth]{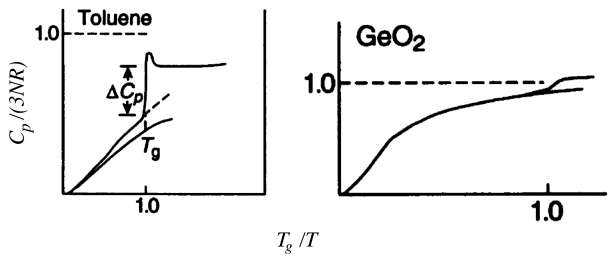}
  \caption{\label{cp_jump} Shown are the temperature dependences of
    the heat capacity for two substances, both for the glassy and
    crystalline counterparts. This figure is a fragment of Fig.2 from
    Ref.\cite{Angell_PNAS}.}
\end{figure}

\begin{figure}[t]
  \includegraphics[width=.75\figurewidth]{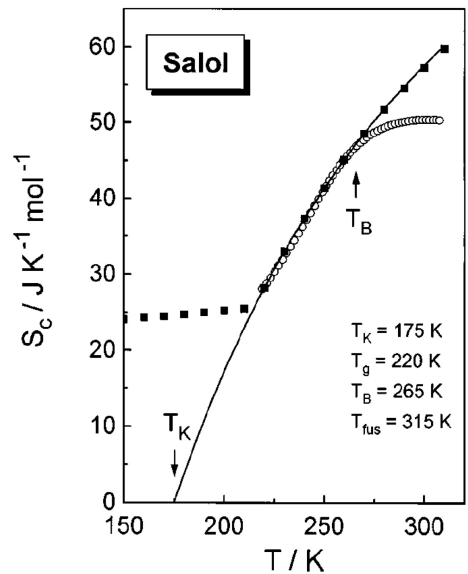}
  \caption{\label{s_c} This figure, from Ref.\cite{RichertAngell},
    illustrates the extrapolation of the experimentally determined
    configurational entropy (shown by squares) below the dynamic glass
    transition temperature $T_g$. This graph illustrates that the
    Kauzmann temperature $T_K$ is equal to the temperature $T_0$ from
    Eq.(\ref{VF}), at which the relaxation times would strictly
    diverge. The circles represent $1/\ln(\tau)$, times a convenient
    scaling constant.}
\end{figure}

In parallel with the dynamical changes upon supercooling, there are
apparent thermodynamic changes. When the system falls out of
equilibrium, thermodynamic susceptibilities nearly discontinuously
decrease.  The most interesting of these is the heat capacity shown in
Fig.\ref{cp_jump}.
The heat capacity allows one to monitor the entropy of liquid
configurations of a supercooled fluid. This discontinuity in the heat
capacity is well approximated by subtracting from the measured heat
capacity of the liquid the heat capacity of the corresponding
crystal. By integrating in temperature, one obtains the part of the
entropy that comes from the diversity of liquid configurations, see
Fig.\ref{s_c}. This excess ``configurational'' entropy is typically of
the order of a few $k_B$ per rigid molecular unit (such as a SiO$_4$
tetrahedron in silica, or an aromatic ring in TNB).

When extrapolated below the glass transition temperature $T_g$, the
configurational entropy, $s_c$, would appear to vanish at a
temperature $T_K$ \cite{Kauzmann}, called the Kauzmann
temperature. Given the magnitude of the heat capacity jump $\Delta
c_p$, at the glass transition, various ways to fit the temperature
dependence of the configurational entropy can be proposed, such as
\cite{RichertAngell}:
\begin{equation} \label{scT} s_c = \Delta c_p (1- T_K/T).
\end{equation}
While there have been disputes about precisely how $s_c$ should be
extrapolated \cite{MArtinezAngell}, it is fairly clear that the
dynamic $T_0$ is equal to the thermodynamic $T_K$ for all
glassformers, see Fig.\ref{s_c}. That the missing translational motion
is the main contributor to the entropy loss is explicitly confirmed by
the frequency dependent heat capacity measurements of Nagel and
coworkers \cite{DixonNagel}, which show the process contributing to
the bulk calorimetry occurs again exactly at the time scales implied
by kinetic measurements, such as the frequency dependent relaxations,
mechanical or otherwise. That this is a time scale for microscopic
movements is also evidenced by the ``plateau'' in time-resolved
neutron scattering, which confirms the local molecular environment
rearranges on the same time scale \cite{MezeiRussina}, see also
Fig.\ref{NSplateau}.

Rather modest deviations from the consonance of time scales of
molecular motions have excited much attention--these are often
discussed as ``decoupling.'' For example, the diffusion coefficient
differs from its Stokes-Einstein value by about two orders of
magnitude at $T_g$. This is two orders of magnitude out of 14. Leaving
aside these modest effects, a staggering amount of data acquired
during the past century show the dramatic slowing is essentially
shared by nearly all motions.

\section{Classical Theory of the Glass Transition and Supercooled
  Liquids}
\label{RFOT}

\subsection{Emergence of a Free Energy Landscape of Aperiodic
  Structural States }

\begin{figure}[t]
  \includegraphics[width=.95\figurewidth]{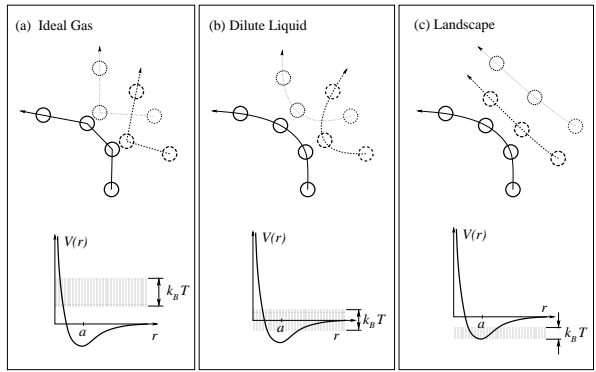}
  \caption{\label{flights} The three panels demonstrate the three
    relatively distinct kinetic regimes of an equilibrium liquid, or a
    conditionally equilibrium liquid, as the case would be below the
    melting temperature $T_m$. See text for explanation. The bottom
    portions illustrate the region of the two-particle potentials
    explored in the corresponding regimes.}
\end{figure}

In gases, molecules spend relatively little time near each other,
moving in straight lines separated by collision events, as illustrated
in Fig.\ref{flights}(a).  In the denser, colder liquid state, the
collision duration grows while the time between collisions becomes
smaller.  These times become comparable in the liquid
state. Nevertheless, any two molecules are unlikely to stay together
for any significant time. In other words, consecutive collisions near
the critical point usually, but not always, occur between different
pairs, as in Fig.\ref{flights}(b). The high temperature liquid is just
a dense gas, dynamically.

\begin{figure}[t]
  \includegraphics[width=.85\figurewidth]{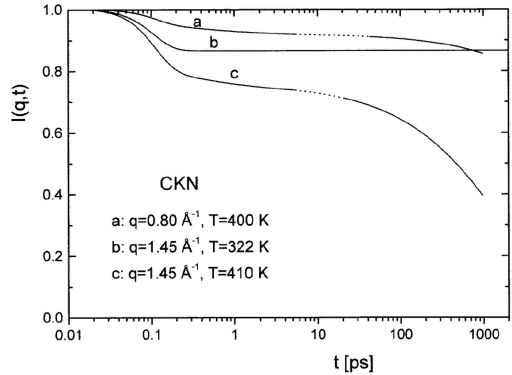}
  \caption{\label{NSplateau} The plateaus in time resolved neutron
    scattering structure function provide a direct proof of long-lived
    local structure in supercooled liquids. The curve (b) was obtained
    below the glass transition. The very flat plateau implies there is
    only a relatively small degree of structural rearrangements in the
    frozen lattice.  This figure is taken from
    Ref.\cite{MezeiRussina}:}
\end{figure}

As temperature decreases and density increases still further, groups
of consecutive collisions will occur ever more often within nearly the
same set of molecules, see Fig.\ref{flights}(c). Each molecule begins
to reside within a specific ``cage'' for a discernible
time. Alternatively, we may say persistent local liquid structures
form at these temperatures. Perturbative expansions, that keep track
of the very many collision sequences within local regions, require
computing increasingly higher order correlation functions
\cite{Dorfman}. At sufficiently high density, it is convenient to
change one's perspective of liquid state dynamics from considering the
dynamics of a dense gas to motions within sets of aperiodic crystal
structures. One notices that even though each long-living structure
involves many correlated many-particle events, each particle within a
structure is most often doing something quite simple. It is vibrating
about a fixed location and only occasionally moves to a new
location. This is an experimental fact, as evidenced by neutron
scattering data which exhibit a plateau in the time dependent
structure factor, see Fig.\ref{NSplateau}.

The motion of a particle is ever more confined to occur within the
cage, so individual molecular bonds distort very little, allowing one
to treat these persistent motions as approximately harmonic at high
enough densities. When a particle takes up a new position of
residence, this implies it has moved beyond a certain threshold
distance (see below). This threshold is reminiscent of the Lindemann
melting criterion \cite{Lindemann} for periodic crystals.  When a
particle takes up a new residence position it usually does not do so
alone. Individual bond breaking and vacancy formation are just as rare
processes in a deeply supercooled liquid as they are in crystals,
which have comparable local density and stiffness. Instead, usually a
group of particles moves whenever a net displacement of an individual
molecule occurs.

\begin{figure}[t]
\includegraphics[width=0.45 \figurewidth]{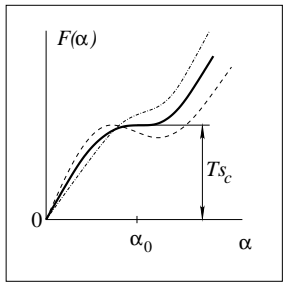}
\caption{\label{F_alpha} This is a schematic graph of the free energy
  density of an aperiodic lattice as a function of the effective
  Einstein oscillator force constant $\alpha$ ($\alpha$ is also an
  inverse square of the localization length used as input in the
  density functional of the liquid. Specifically, the cureves shown
  characterize the system near the dynamical transition at $T_A$, when
  a secondary, metastable minimum in $F(\alpha)$ begins to appear as
  the temperature is lowered.}
\end{figure}

The mostly harmonic nature of individual displacements in long-living
structures allows one to use a simple theory to describe the emergence
of aperiodic crystals using density functionals \cite{dens_F1,
  dens_F2}. In equilibrium statistical mechanics the free energy can
be written as a functional of a non-uniform density. There is an
entropic cost for forming such a non-uniform density, but if particles
are localized, they can avoid each other to compensate. By combining
the ideal gas localization entropy with an effective interaction,
Ramakrishnan and Yussouff \cite{RamakrishnanYussouff, Yussouff} wrote:
\begin{widetext}
\begin{equation} 
  F[\rho({\bf r})] = k_B T \int d^{3}{\bf r} \rho({\bf r}) [\ln \rho({\bf
    r})-1] + \frac{1}{2} \int \int d^{3}{\bf r} d^{3}{\bf r}' [\rho({\bf
    r}) - \rho_0] c({\bf r}, {\bf r}'; \rho_0) [\rho({\bf r}') -
  \rho_0] + F_{\suni},
\label{F_rho}
\end{equation} 
\end{widetext}
where $F_{\suni}$ is the free energy of completely uniform
liquid. More general functionals also have been constructed and used
\cite{Haymet}.  Crystallization would be described by a periodic
$\rho({\bf r})$, but one can also use such functionals to examine the
stability of any conceivable state with persistent {\em aperiodic}
density.  In view of the mostly harmonic nature of individual cages,
we can employ the following variational density profile to dress any
mechanically stable zero temperature structure:
\begin{equation} \label{rho_alpha}
  \rho({\bf r}) \equiv \rho({\bf r},\{{\bf r}_i\}) =
  \sum_i \left(\frac{\alpha}{\pi} \right)^{3/2} e^{-\alpha ({\bf r}-{\bf
      r}_i)^2}.
\end{equation}
Here $\{{\bf r}_i\}$ represent the atomic coordinates of a particular,
but generic aperiodic lattice characterized by some average density $n
\equiv 1/a^3$, where $a$ is the average lattice spacing. The
variational parameter $\alpha$ is a convenient measure of localization
within such a many-particle cage. While nonzero $\alpha$ characterizes
a localized regime, where transient local quasiharmonic environment
forms, the same function is also capable of describing the completely
delocalized regime of the uniform liquid by taking $\alpha=0$. Singh,
Stoessel, and Wolynes \cite{dens_F2} established that at sufficiently
high density, at a given temperature called $T_A$, the free energy
from Eq.(\ref{F_rho}) develops a metastable minimum as a function of
$\alpha$, see Fig.\ref{F_alpha}. The transition is very similar to a
spinodal crystallization, but to an aperiodic structure. Similar
results were obtained earlier by Stoessel and Wolynes \cite{dens_F1}
using a self-consistent phonon approach. The density functional and
self-consistent phonon approaches both suggest that at $T_A$,
metastable structures form. This metastability indicates there would
be a corresponding barrier demarcating one metastable minimum in the
entire set of such minima from the uniform liquid state. The emergence
of the minimum appears as a first order transition, where $\alpha$
plays the role of an order parameter, whose value changes
discontinuously, during the transition, from $\alpha=0$ to
$\alpha=\alpha_0$.

The quantity $1/\sqrt{\alpha_0}$ measures the extent of vibrational
motion within a local metastable structure. Motions of any greater
extent, at the mean-field level, signify a transition between
thermodynamically distinct states. The quantity $1/\sqrt{\alpha_0}$ is
the vibrational amplitude at the mechanical stability edge, fully
analogous to the famous Lindemann length for ordinary crystals
\cite{Lindemann}. We will denote it as $d_L$:
\begin{equation}
  d_L  = 1/\sqrt{\alpha_0}.  
\end{equation}
The Lindemann ratio $d_L/a$ is roughly one-tenth for all crystals
\cite{Gilvarry, Bilgram, GuptaSharma}, and a similar value emerges
from the microscopic calculations for aperiodic structures. This value
is also consistent with the magnitude of the plateau in the neutron
scattering time correlation function \cite{MezeiRussina}. While in
terms of $\alpha_0$ the transition is first order, there are many
aperiodic states randomly distributed in free energy. We therefore
call this transition the ``Random First Order Transition'' (RFOT)
because, while first order in $\alpha$, it results in forming many
random (infinite lifetime) ``phases'' which are distinct both
morphologically and spatially. Because of the multiplicity of states
there is no latent heat but instead at the mean field level, there
would be a heat capacity discontinuity much as occurs when a liquid
falls out of equilibrium.

Even though the metastable minimum of $F(\alpha)$ is higher in free
energy than the uniform liquid state, we realize that the $F(\alpha)$
curve was computed for a {\em single} (typical) aperiodic lattice. The
full liquid free energy available at $\alpha_0$ incorporates the
multiplicity of {\em all} such (typical) aperiodic states,
i.e. $e^{s_c N}$ for a region encompassing $N$ molecules.  The
multiplicity of aperiodic states at the ``spinodal'' temperature $T_A$
obviously must imply the localized minimum is metastable. Whether at a
sufficiently low temperature it can reach $F_\suni$ and thus imply the
existence of an observable thermodynamic state, is not immediately
obvious. Yet several rigorous arguments suggest this is indeed the
case. Kirkpatrick and Wolynes \cite{MCT} showed that the mode-coupling
viscosity catastrophe, which can be obtained by summing recollision
events, also corresponds to the same transition at $T_A$ predicted by
self-consistent phonon and variational approaches. They noticed the
connection with Potts spin glasses. Kirkpatrick and Thirumalai
clarified this relation which KW had explicitly showed only in high
dimension for fluids \cite{KT_PRL87, KT_PRB87}. Like the approximate
treatment of the fluid system, the exactly soluble mean-field Potts
glass also exhibits a dynamic transition at a temperature $T_A$ above
its thermodynamic glass transition.  Kirkpatrick and Wolynes
\cite{MCT1} went on to show that these infinite range Potts spin
models, too, exhibited a Kauzmann-like entropy crisis at the lower
thermodynamic temperature $T_K$. They also suggested that for finite
range Potts spin glasses and supercooled liquids the $T_A$ transition
would be smeared by droplet-like excitations driven by the
configurational entropy. These ``entropic droplet'' excitations would
provide the route to equilibration below the mean field dynamical
temperature that corresponds to the mode coupling transition. Their
argument also explained why the dynamic $T_0$ and $T_K$ should be the
same. These entropic droplet excitations also form the basis of the
microscopic developments of RFOT theory that in recent years, have
explained supercooled liquid and glasses quantitatively. We will
describe entropic droplets in a somewhat different language in the
following section.

The said mean field picture, which employed density functional ideas
to treat the fluid aspects and the analogy to the exactly solvable
Potts spin glasses, has been developed more formally using the methods
of replica field theory. By making reference to the molecular fluid
configuration at one time as a fiducial structure, the replica method
can be used to construct stability criteria and develop approximations
for the configurational entropy directly \cite{repl_Lind}. Numerous
analyses using these tools confirm the earlier mean field
developments. These more explicit methods have also been tested near
the dynamic transition by computer simulations \cite{Coluzzi}.  The
reformulation of the theory using replicas also has allowed the
activated events to be studied in a systematic way when the range of
the interactions is finite but large \cite{FranzHertz,
  SchmalianWolynes2000, BouchaudBiroli}. These mathematically
controlled developments increase the confidence in the simpler
constructive arguments for structural glasses that we will describe
below.  Another direction in which the microscopic calculations can be
taken is to explore the molecular origins of the configurational
entropy. Hall and Wolynes recently used these approaches to explain
why network materials with more constraints (like SiO$_2$) behave like
``strong'' liquids while simple van der Waals systems without bonding
constraints are more fragile \cite{HallWolynes}.

It is important to note that within the RFOT theory, detailed
microscopic calculations starting from the intermolecular forces can
in fact be carried out. Presently, only the mode coupling theory can
make a comparable claim. Mode coupling theory (MCT) is not an
orthogonal approach, however. As Kirkpatrick and Wolynes pointed out
in 1987 \cite{MCT} the two approaches are effectively equivalent near
the dynamical crossover. Recent successes of MCT in describing
re-entrant glass transitions in attractive colloids, etc. thus
buttress the general RFOT picture from a microscopic viewpoint.
Microscopic calculations set the stage for the RFOT theory. Much as in
thinking about conventional phase transistion, it is important,
however, to distinguish the main RFOT ideas from the liquid state
engineering details for specific systems. It is clear, for example,
that the deep supercooled regime of van der Waals liquids probes
intermolecular forces at a more intimate length scale than near the
critical point. Time-honored prescriptions based on perfectly hard
potentials thus will break down. This breakdown is most clearly
evidenced by the experimental fact that the isochoric activation
energy of the viscosity is not zero below $T_A$, while it is nearly so
above that point. Just as for studies of crystals \cite{Ree}, the
popular Weeks-Chandler-Andersen prescription
\cite{WeeksChandlerAndersen}, for dividing the potential, requires
re-examination in deeply supercooled liquids.

\subsection{The Library of Local States}
\label{RFOT_section}

In mean field theory, the lifetime of an aperiodic structure is
infinite. But in reality the lifetime of a system with finite range interactions is
finite because rearrangements can occur independently within regions
of finite spatial extent. The number of relaxation events per unit
time scales with the system size. A mole of water will undergo roughly
$10^{39}$ barrier crossing events a second. These events are local in
character in that they affect each other little beyond some critical
distance.

The locality of structural processes in supercooled liquids is
intrinsic in the entropic droplet concept. Recently their quantitative
aspects have been constructively established within the RFOT theory
\cite{KTW, XW}. Our present discussion will be based on the
``library'' construction of local configurations \cite{LW_aging},
which makes the ideas more transparent than in earlier discussions and
also allows the treatment of the ``aging'' regime. This library
construction is essential for the latter far-from-equilibrium case.
An elegant formal analysis along similar lines has been made by
Bouchaud and Biroli \cite{BouchaudBiroli}. This construction can also
be quantized to deal with the phenomena of cryogenic temperatures
\cite{LW}. In the library construction, one first averages over the
vibrational modes of the supercooled liquid. These equilibrate
generically in a time less than a picosecond. The variational ansatz
used in density functional theory from Eq.(\ref{F_rho}) operationally
defines such an averaging. The lifetimes of the resulting aperiodic
crystal states depend on the size of the region. For the library
construction to be strictly valid, the processes must be slower than
the sound modes of similar length by three orders of magnitude or
so. Fortunately, this criterion covers most of the dynamical range
accessed by supercooled melts, except perhaps for the high frequency
processes referred to as the Boson peak in liquids
\cite{LW_BP}. Transitions between the aperiodic crystal states defined
above give rise to the configurational entropy.

\begin{figure}[t]
 \includegraphics[width=\figurewidth]{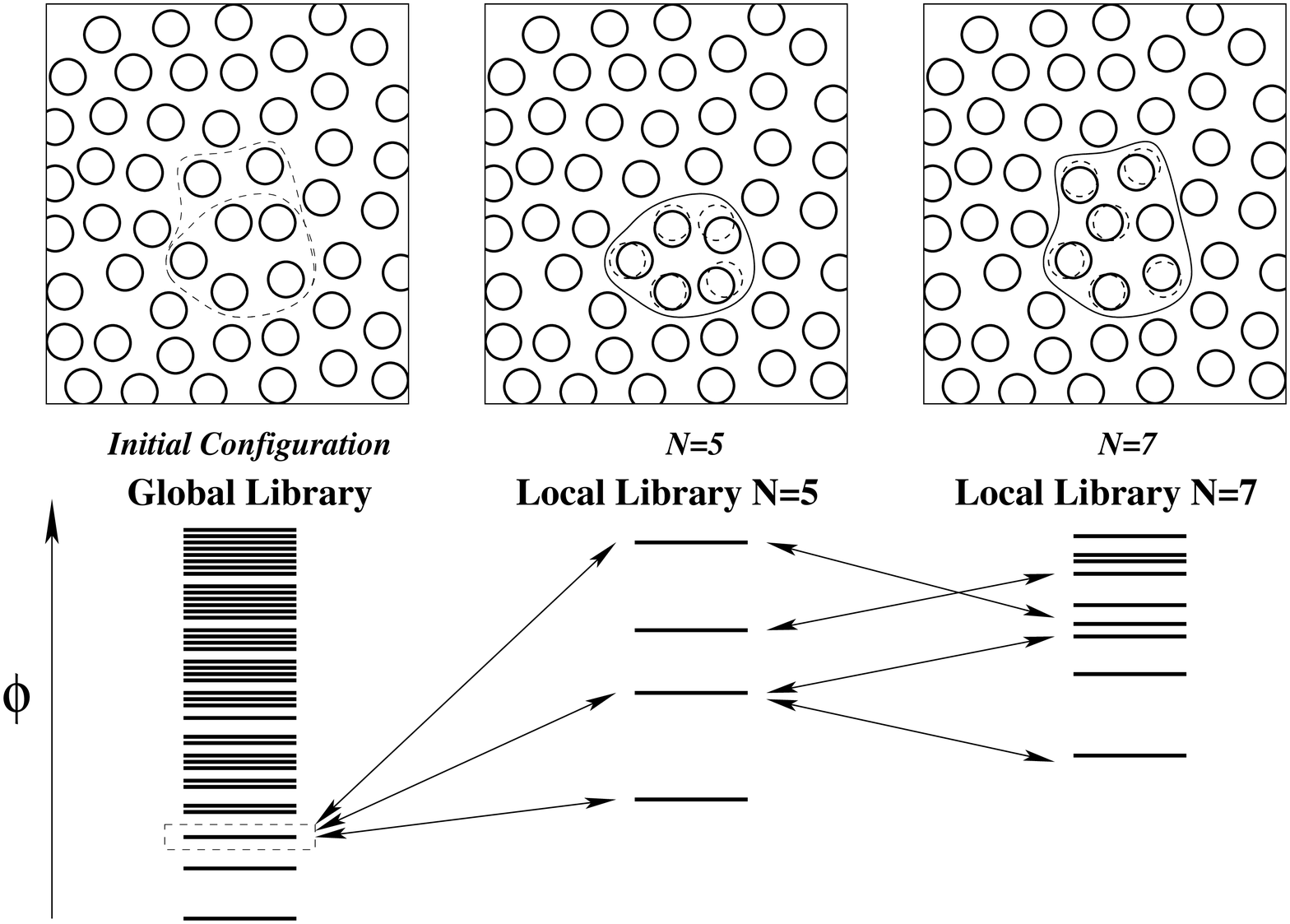}
 \caption{\label{library} This figure is taken from
   \cite{LW_aging}. In the upper panel on the left a global
   configuration is shown, chosen out of a global energy landscape. A
   region of $N=5$ particles in this configuration is rearranged in
   the center illustration; this subset is taken from another
   aperiodic state or from a different location of the same liquid
   state. The original particle positions are indicated with dashed
   lines. A larger rearranged region involving $N=7$ particles is
   connected dynamically to these states and is shown on the right. In
   the lower panel, the left most figure shows the huge density of
   states that is possible initially. The density of states found in
   the local library originating from a given initial state with 5
   particles being allowed to move locally is shown in the second
   diagram. These energies are generally higher than the original
   state owing to the mismatch between the two structures. The larger
   density of states where 7 particles are allowed to move is shown in
   the right most part of this panel. As the library grows in size,
   the states as a whole are still found at higher energies but the
   width of the distribution grows.  Eventually with growing $N$, a
   state within thermal reach of the initial state will be found.  At
   this value of $N^*$ we expect a region to be able to equilibrate. }
\end{figure}

\begin{figure}[t]
  \includegraphics[width=.95\figurewidth]{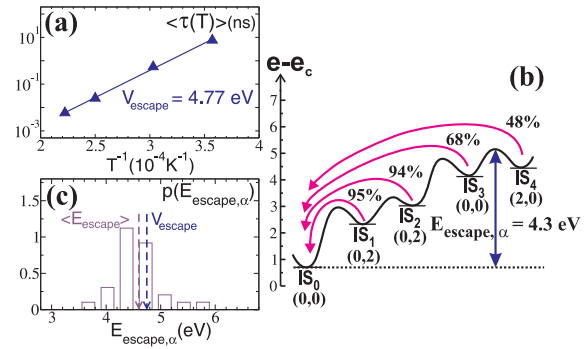}
  \caption{\label{steps} Even in ``strong'' liqiuds, reconfigurations
    occur through a series of steps as in the library
    constuction. Saksaengwijit and Heuer have delineated these events
    in silica. Their results are shown with their Fig.4 \cite{SHeuer}.
    Panel (a) shows the temperature dependence of the average waiting
    time in the low energy inherent structures of simulated silica.
    In (b), a specific escape trajectory from an inherent structure is
    represented.  The percentages indicate how often such a structure
    recurs to its initial state. Panel (c) gives the distribution of
    escape barriers from IS.}
\end{figure}

To convert from one state to another state we must consider not only
direct paths connecting one structure to a neighboring one; but most
importantly, one must examine all thermally realizable dynamically
connected sequences containing many structural states. By
``dynamically connected'' we mean the trajectory is a sequence of
processes with small barriers that involves a translation of a single
bead by a distance not exceeding the Lindemann length $d_L$. The
lowest energy path leading to a thermally representative liquid state
will define the most probably escape trajectory from the initial
state. The highest free energy point along the path will determine the
bottleneck, or the critical value of the progress coordinate.  The
energy at the bottleneck will determine the barrier height and will
thus give the activation part of the escape rate. A very nice example
of a computational study of such a sequence is provided by
Saksaengwijit and Heuer \cite{SHeuer}. They show that even in high
temperature silica, whose rates are nearly Arrhenius, reconfiguration
events proceed through several steps, see Fig.\ref{steps}.

Now, the totality of the aperiodic states that could be constructed
with a region of fixed boundaries forms a ``library'' of local
states. This construction is illustrated in Fig.\ref{library}. In this
figure, the circles do not signify the instantaneous positions of the
molecules but rather their most probably positions after vibrational
averaging. Further, the circles correspond to effective individual
structural units, or ``beads''.  Chemical intuition often makes clear
how to decompose conceptually a molecule into beads. A rigid chemical
group, a single benzene ring, or a compact side-chain in a polymer
will usually constitute an independently moving unit. The volumetric
bead density can be quantified unambiguously by comparing with the
fusion entropy, if the corresponding crystal exists, since bead
motions are also frozen in the crystal; see the detailed discussion in
\cite{LW_soft}.  At any rate, a bead is usually a few angstroms in
size.

Consider an initial configuration spanning some extended region. Its
``intrinsic'' bulk energy can be written:
\begin{equation} \label{phiin} \Phi^{\sbulk}_{in} = E_{in} - T
  S_{\svibr},
\end{equation}
which combines the energy proper of the particular arrangement of the
most probable molecular coordinates with the free energy of the high
frequency vibrations.

This state may move to some other thermally representative aperiodic crystal
state over the same region, labelled by a dummy index $j$ whose bulk
energy is:
\begin{equation} \label{phij} \Phi^{\sbulk}_j = E_j - T S_{\svibr}.
\end{equation}

Choose a boundary $\cB$ encompassing $N$ beads somewhere within the
initial structure, and remove its contents. We then cut an identically
shaped region out of the structure $j$ and paste this region in the
void in the initial structure. Call the modified region of the initial
state a ``droplet''.  The free energy of this trial configuration is:
\begin{equation} \label{phi_phi} \phi_{j} - \phi_{in} =
  \Phi^{\sbulk}_{j} - \Phi^{\sbulk}_{in} + \Gamma_{j, in}.
  \vspace{1mm}
\end{equation}
This free energy includes a mismatch penalty $\Gamma_{j, in}$ because
no matter how carefully we placed the surface $\cB$ in the structure
$j$, the contents of the latter region were vibrationally averaged
with a constrained surrounding different from the one in the initial
state and thus some ``elastic energy'' must be paid. The alternative
state inside the droplet may be thought of as the original molecules
somewhat displaced from their original positions.  The cut-and-paste
routine realizes a finite number of thermodynamically distinct droplet
states. This finite number is given by the configurational entropy,
and scales exponentially with the droplet size: $e^{s_c N}$.  In the
central panel of Fig.\ref{library}, we show, for illustration, just
the subset of droplet states available at size $N=5$. This subset
would constitute an astronomically minute fraction of the whole set of
states available to an extended sample. Now, consider the droplet
states at different, but close sizes. Some states at both sizes will
be structurally very close, implying high degree of dynamical
connectivity. Sequences of such dynamically connected droplet states
constitute escape trajectories from a given initial extended state to
some other extended aperiodic crystal state. Because one may associate
with any of those sequences of droplet states a sequence of droplet
{\em sizes}, one may therefore speak of {\em nucleation} of one
aperiodic crystal state within another and of the corresponding
propagation of a {\em domain wall} separating those two aperiodic
crystal states.  In this way, the mismatch penalty may be thought of
as the domain wall tension.

Because of the surface penalty, the lowest energy, per particle, in a
droplet library is likely to be well above the energy, per particle,
in the initial extended state. Still, the energies within such finite
subsets are always distributed, roughly in a Gaussian fashion. The
{\em lowest} energy in the subset, it turns out, is determined by a
competition between two factors: On the one hand, with the growing
droplet size, the mismatch penalty will increase thus shifting the
bulk of the distribution of the subset energies upwards. On the other
hand, the weight of the distribution, i.e. $e^{s_c N}$, increases
exponentially with $N$, so that the lowest energy state will be found
at an increasingly lower position. Eventually, at large enough droplet
size $N^*$, the latter trend will compensate the former, so that the
droplet will be guaranteed to have the energy per particle equal to
that of an arbitrarily extended state. Thus to a region of size $N^*$
or larger, all thermodynamically relevant {\em bulk} liquid states
will be available; i.e.  such a region is expected to be typical of
equilibrium at the temperature $T$.  A supercooled liquid is a mosaic
of aperiodic crystals!  In the following we will put this qualitative
discussion in formal, quantitative terms.

{\bf \small Mismatch Penalty between Aperiodic Crystal States.}  To a
first approximation from the density functional viewpoint, computing
the mismatch energy is simply a matter of counting up the missing
interactions from some unsatisfied local contacts in an interface
region. The mismatch penalty would thus scale with the interface area
itself, times the energy of the unsatisfied bond: $\Gamma \propto
\epsilon r^{d-1} l_\smicro$, where $\epsilon$ and $r$ are the
interface energy density and the region size respectively \cite{MCT1}.
This is also the result of instanton calculations based on replica
methods \cite{Franz, DSW2005}. However, the situation is a bit more
complicated as we must recognize that some of the states in the
library will in fact match much better than do others because they are
partially random. The scaling of thermal averaged $\Gamma$ with $N$
will thus be weaker than expected for interfaces between entirely
distinct phases. The fraction of the better matching configurations is
significant enough to actually partially short-circuit the
conventional surface tension, renormalizing it to a smaller
value. From an elastic theory point of view, we expect the
displacement fields due to the local structural rearrangements in our
aperiodic structures not to exhibit long range
correlations. Accordingly, the deviation of total strain within a
region of size $M$ from its average value would scale with the usual
$\sqrt{M}$, and is of either sign with equal probability. This means
one may generically impose a random external field with the usual
Gaussian statistics to lower the energy of a region of size $M$ by an
amount scaling with $\sqrt{M}$. A curved interface between two states
will distort so as to lower the local free energy to take advantage of
this random contribution. The energy compensation will scale, again,
as the square root of the variation of the volume occupied by either
of the two phases, due to the boundary distortion. The final shape of
the interface will be determined by a competition between this
stochastic energy compensation and the cost of increasing the area of
a flat interface. Consistent with this, the scaling of surface tension
with $r$ differs from the conventional $r^{(d-1)/d}$. This situation
is analogous to the problem of the interface between the spin-up and
spin-down regions in the Random Field Ising Model (RFIM), which has
been treated by Villain \cite{Villain}. This analogy was explicitly
exploited by Kirkpatrick, Thirumalai, and Wolynes (KTW) \cite{KTW} in
deducing how the droplet interface tension scales with the droplet
size. (One may in fact show that the two problems can be mapped onto
each other, with the error not exceeding the discrepancy between the
microcanonical and canonical averages \cite{Lubchenko_unp}.) A
pleasant dividend of this argument is that the hyperscaling relation
connecting the heat capacity and the mosaic length scaling near $T_K$:
$(2-\alpha = \nu d)$, - is restored \cite{KTW}. The KTW argument
suggests the effective surface tension coefficient is renormalized
according to:
\begin{equation} \label{sr} \sigma(r) \sim \sigma_0 (r/a)^{-(d-2)/2},
\end{equation}
where $\sigma_0$ is the surface tension coefficient at the molecular
length scale.  In all spatial dimensions $d > 2$, the scaling of the
resulting mismatch free energy
\begin{equation} \label{d/2}
  \Gamma \propto r^{d/2}
\end{equation}
is numerically inferior to the conventional $r^{(d-1)/d}$ scaling, but
of course this is quantitative only for sufficiently large interfaces.
It also follows that the thickness of the distorted interface scales
with the radius itself \cite{Villain,KTW} consistent with our {\it
  \`{a} priori} expectation that the mismatch energy is determined by
the subsets of static structures that smoothly interpolate between two
generic extended aperiodic states. It may be said that an interface
between two thermally averaged aperiodic crystals is, strictly
speaking, not thin but is always ``wetted'' by other states.

On short length scales, the coefficient $\sigma_0$ in Eq.(\ref{sr})
must correspond to the mismatch penalty at the molecular length
scale. This suggests a way in which $\sigma_0$ can be estimated in
zeroth order, at the molecular length-scale. The mismatch requires the
particle at the interface to still be within the Einstein oscillator
localization volume $d_L^3$, instead of its volumetrically available
space $a^3$. Yet in the free energy functional it does not receive the
full benefit of its neighbors' being localized (only half of them are)
and thus staying out of the way. Since these free energies must
balance at $T_K$, there is an additional mismatch penalty on this
scale that is related to the localization entropy $\sigma_0 \sim
\frac{1}{2} k_B T \ln(a^3/d_L^3) = (3/4) k_BT \ln(a^2/d_L^2)$. This
expression for $\sigma_0$ contains all the essential parameter
scaling, but obviously there may be some numerical uncertainty in
applying the long distance scaling of $\sigma(r)$ all the way down to
the shortest lengths. Nevertheless, by virtue of the slow, logarithmic
dependence on the constants, along with the near universal value of
the Lindemann ratio $d_l/a = 0.1$, this argument suggests $\sigma_0$
is a universal multiple of the glass transition temperature
itself. This is the key to the later quantitative results obtained in
the RFOT theory. Using the density functional approximation of
Eqs.(\ref{F_rho}) and (\ref{rho_alpha}), Xia and Wolynes specifically
obtained the surface tension at the molecular scale \cite{XW}:
\begin{equation} \label{sigma0}
  \sigma_0 = \frac{3}{4} \frac{k_B T}{a^2} \ln[(a/d_L)^2/\pi e],
  \mbox{  in } \sigma(r) = \sigma_0 (a/r)^{1/2}.
\end{equation}
This value works very well quantitatively for real materials. The
chemical universality of $\sigma_0$ reminds one of Turnbull's rule
used in treating crystallization that empirically states the tension
between a periodic crystal and its melt is a universal multiple of
$T_m$ \cite{Turnbull}. It is convenient to write the mismatch free
energy in terms of the number of reconfigured particles: $N \equiv
(4\pi/3) (r/a)^3$:
\begin{equation} \label{gamma} \Gamma = \gamma \sqrt{N}, \mbox{ where
  } \gamma \equiv \frac{2\sqrt{3\pi}}{2} k_B T
  \ln\left[\frac{(a/d_L)^2}{\pi e}\right].
\end{equation}
While the mismatch energy $\Gamma_{in, j}$ is doubtless distributed,
as long as we are dealing with typical liquid states, we can use its
typical value $\Gamma$.

\subsection{Activated Motions between Local States}

We see, via the library construction, that if too small a region is
reconfigured, even the lower energy paths will have a monotonically
increasing energy with the size. But as the region is made larger,
some fraction of paths will curve down in energy. At sufficiently
large size, there will always be a trajectory ending at an energy
within the thermally relevant liquid energy range.  In computing the
typical escape rates from a liquid state, one therefore only needs to
escape by passing through droplet configurations with the size
corresponding to the lowest flux. Formally averaging the escape flux
over the ensemble of transition state droplets yields \cite{LW_aging}:
\begin{eqnarray} 
  k & = & \tau^{-1}_{0} \int (d \phi_j/c_\phi)
  e^{S_c(\Phi^{\tbulk}_j)/k_B} e^{-(\phi_j-\phi_{in})/k_B T}
  \nonumber \\ & \simeq & \tau^{-1}_{0}
  e^{S_c(\Phi^{\tbulk}_{opt})/k_B} e^{-(\phi_{opt}-\phi_{in}^{lib})/k_B T}
  \nonumber \\ & \equiv & \tau^{-1}_{0}
  e^{S_c(\Phi^{\tbulk}_{eq})/k_B} e^{-(\phi_{eq}-\phi_{in}^{lib})/k_B T}.
\label{k}
\end{eqnarray}
The intrinsic activation free energy represents a critical
droplet-configuration energy from Eq.(\ref{phi_phi}) and the factor
$e^{S_c(\Phi^{\tbulk}_j)/k_B} $ gives the multiplicity of liquid
configurations at the energy $\phi_j$. ($c_\phi$ is a normalization
constant.) The energy $\phi_{opt}$, maximizing the integral, must be
assigned the equilibrium energy value.  Consequently, the log of the
multiplicity, $S_c(\Phi^{\sbulk}_{eq})$, is nothing but the
configurational entropy at equilibrium, which can be measured by
calorimetry: $S_c(\Phi^{\sbulk}_{eq}) = S_c$.

The $N$ dependence of all the parameters in Eq.(\ref{k}) becomess
obvious from Eq.(\ref{phi_phi}) upon recalling that $S_c$ and
$\Phi^{\sbulk}$ are bulk parameters and scale with $N$ itself: $S_c =
N s_c$, $\Phi^{\sbulk} = N \phi^{\sbulk}$.  One thus obtains a simple
expression for the {\em typical} free energy profile during a
structural rearrangement after optimizing with respect to the saddle
point energy. It varies with size as:
\begin{equation} \label{F(N)} F(N) = [\phi_{eq}^\sbulk(T) -
  \phi_{in}^\sbulk(T)] N + \gamma \sqrt{N} - T s_c N.
\end{equation}
Here, we have used Eq.(\ref{phi_phi}). $F$ does reach down to
arbitrarily low energy states, but to escape, one needs to pass first
over the maximum, which gives the typical barrier $F^\ddagger$ for
reconfiguration from a given initial state. When $F= 0$, a typical
state has already been reached.

Consider supercooled liquids equilibrated above the glass transition
temperature $T_g$. The initial liquid state is thermodynamically
typical of the temperature $T$ thus $\phi_{in}^\sbulk(T) =
\phi_{eq}^\sbulk(T)$. The resulting nucleation profile is quite
simple:
\begin{equation} \label{F(N)eq} F(N)|_{T > T_g} = \gamma \sqrt{N} - T
  s_c N.
\end{equation}
Clearly, structural transitions are driven by configurational entropy alone!
Eq.(\ref{F(N)eq}) immediately gives an inverse scaling of the most
probable relaxation barrier with the configurational entropy density that
automatically yields the Vogel-Fulcher law:
\begin{equation} \label{Fsc} F^\ddagger = \frac{\gamma^2}{4 s_c T} =
  \frac{\gamma^2}{4 \Delta c_p (T-T_K)},
\end{equation}
Here we used the specific form (\ref{scT}) for the configurational
entropy.  Note that if it were not for the surface tension
renormalization, see Eq.(\ref{d/2}), the $s_c$ dependence in the
denominator of the middle expression would be quadratic, not linear.
The inverse scaling of the relaxation barrier with the configurational
entropy was derived by KTW \cite{KTW}, but was proposed originally by
Adam and Gibbs (AG) \cite{AdamGibbs}, who stipulated that there be a
smallest rearranging unit in a liquid. This AG unit would be
characterized by two configurations, whose size was assumed
independent of temperature. The Adam-Gibbs argument does not reconcile
how the existence of such a special fixed size is compatible with that
argument's use of a temperature dependent, {\em extensive}
configurational entropy.  In contrast, within RFOT theory the length
scale of activation is determined by the underlying Hamiltonian and
varies with temperature. The RFOT thoery precisely predicts the way in
which the critical size of the nucleation barrier scales in
Ref. \cite{KTW}:
\begin{equation} \label{rsc}
  r^\ddagger \propto (N^\ddagger)^{1/3} \propto  \frac{1}{s_c^{2/3}} \propto
  \frac{1}{(T-T_K)^{2/3}}.
\end{equation}

This scaling law is consistent with a specific heat discontinuity and
the usual hyperscaling relation for continuous transitions
\cite{KTW}. RFOT theory directly shows the kinetic and thermodynamic
anomalies in supercooled melts are intrinsically related. Furthermore,
given the value of $\sigma_0$ from Eq.(\ref{sigma0}), Xia and Wolynes
established an amazingly simple relation between the kinetic fragility
and the heat capacity discontinuity:
\begin{equation} \label{32} D = 32./\Delta c_p.
\end{equation}
The universal numerical constant $32.$ comes from the numerical
constants in the argument and the universal value of the Lindemann
ratio. $\Delta c_p$ is the heat capacity jump per bead. Owing to the
universality of the Lindemann ratio, RFOT theory implies the material
dependence of the fragility comes essentially from the heat capacity
only. Since the Lindemann length enters under the logarithm, its small
variations do not affect the constant significantly. In any event, a
recent argument by Lubchenko \cite{L_Lindemann} shows the Lindemann
ratio $d_L/a$ would be expected to vary at most my 10\% in supercooled
liquids.

The present argument assumes the liquid to be sufficiently deeply
supercooled so that deviations from the activated picture are
small. This is true already when relaxations are about three orders of
magnitude slower than the times of vibrational temperature
equilibrium. The activated picture thus applies at $\tau > 10^{-9}$
sec, which covers most of the liquid dynamic range. At higher
temperatures, collisional effects on the viscosity become
important. The agreement of the relation of Eq.(\ref{32}) between $D$
and $\Delta c_p$ with available data is impressive. Nevertheless, to
avoid fitting ambiguities, it appears best to use the measured
fragility index near $T_g$: $m=T[d \log_{10} \tau(T)/d(1/T)]$. The
index $m$ scales roughly inversely proportionally with $D$, but is
less dependent on the fitting ambiguities arising from the crossover
to the high temperature region. Like $D$, $m$ also follows from
$\Delta c_p$ without adjustable parameters. Fig.6 shows how theory and
experiment compare for the predicted $m$ from thermodynamics and the
measured kinetic values.  There are a small number of outliers. In
examining those exceptions one must bear in mind, however, the RFOT
theory applies strictly only to purely amorphous and fully
equilibrated melts. Deviations from the RFOT predictions are expected
when samples exhibit partial crystallinity, often present in polymers,
or other types of local order, as in decalin or in alkali borates (see
e.g. \cite{NovikovDingSokolov}),
\begin{figure}[t]
\vspace{2mm}
\includegraphics[width=.75\figurewidth]{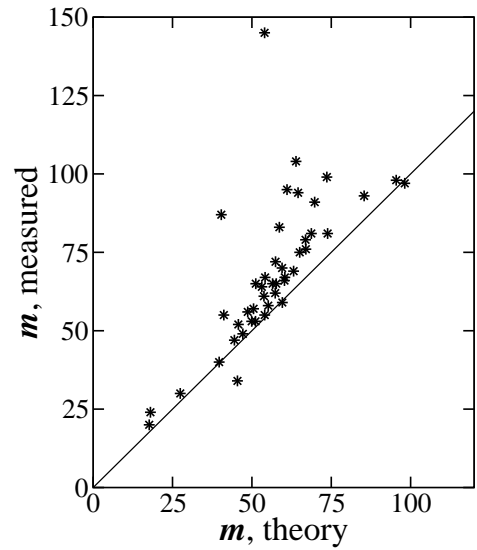}
\caption{\label{jake} The horizontal axis shows the value of fragility
  as computed from the thermodynamics by the RFOT theory, and the
  vertical axis contains the fragility directly measured in kinetics
  experiments. Here $m$ is the so called fragility index, defined
  according to $m=T[d \log_{10} \tau(T)/d(1/T)]$. $m$ is somewhat more
  useful than the fragility $D$, because deviations from the strict
  Vogel-Fulcher law, $\tau = \tau_0 e^{D T_K/(T-T_K)}$, are often
  observed, see text. $m$ essentially gives the apparent activation
  energy of relaxations at $T_g$, in units of $T_g$, it is roughly an
  inverse of $D$.  In evaluating $m$ theoretically, one needs to know
  the size of the moving unit, or ``bead'', in each particular
  liquid. The latter can be estimated using the entropy loss at
  crystallization and scaling it to its Lennard-Jones value
  \cite{LW_soft}, resulting in $m_\stheor \propto \frac{\Delta
    c_p(T_g) T_g}{\Delta H_m s_c^2(T_g)} \propto \frac{\Delta c_p(T_g)
    T_g}{\Delta H_m}$, in view of the near universality of $s_c(T_g)$
  (see text). This figure is taken from \cite{StevensonW}.}
\end{figure}

Whether such local order effects exist may actually be judged, to some
extent, using another simple result that follows from the RFOT theory:
Near $T_g$, relaxations are strictly activated, implying the typical
relaxation barrier relative to the temperature depends only
logarithmically on the relaxation rate: $({F^\ddagger}/{T})_{T_g} =
\ln(\tau/\tau_0)$.  The quenching rate in the laboratory is limited by
the sample's heat conductance, on the faster side, and the
experimenter's patience and stable temperature conditions, on the
longer side. Combining a one hour time scale with Eqs.(\ref{gamma})
and (\ref{Fsc}), RFOT predicts the configurational entropy at $T_g$,
per bead, should be \cite{XW}:
\begin{equation} \label{scTg}
  s_c (T_g) \simeq 0.8,
\end{equation}
depending only logarithmically on the ratio $\tau/\tau_0$. The latter time ratio,
at $T_g$, is very large: $10^{16}$ or so, rendering the estimate above
nearly universal.  A brief look at the configurational entropies of
polymers, such as in Fig.4 of Roland at el. \cite{Polymer_sc}
immediately gives away there are many partially crystalline polymers. For example, PVC
has $s_c(T_g) \simeq .1$ clearly indicating a large degree of
crystallinity, consistent with X-ray determinations of the latter.

Another important universality predicted by the RFOT theory is the
size of the cooperatively rearranging region at the glass transition
temperature. For the droplet size $N > N^*$, where $F(N^*) = 0$, one
expects all thermally relevant liquid states have become available;
there is no further ``growth'' of an alternate phase. Typically then
$N^*$ beads are reconfigured during a typical relaxation event.  Using
\begin{equation}
  N^* \equiv (\xi/a)^3,
\end{equation}
one easily finds \cite{XW}:
\begin{equation} \label{xi_univ} \xi|_{T=T_g} = 5.8 \: a,
\end{equation}
for $\tau/\tau_0 = 10^{17}$, i.e. at the relaxation time of the order
an hour. This cooperativity length scale $\xi$ is not structural, but
dynamic, and hence must be probed by nonlinear dynamic
experiments. Because any region of size $\xi$, but no smaller, may
relax, we can regard a supercooled liquid as a ``mosaic'' \cite{XW} of
cooperatively rearranging regions or ``entropic droplets''.  Like the
critical radius $r^\ddagger$, the dynamical heterogeneity of length
scale $\xi$ scales according to
\begin{equation} \label{xiT} \xi \propto \frac{1}{(T-T_K)^{2/3}}.
\end{equation}

Eqs.(\ref{xi_univ}) and (\ref{xiT}) are arguably the most specific
{\em microscopic} predictions of the RFOT theory; they are certainly
consistent with the earlier semi-quantitative results of the 4D
exchange NMR experiments by Tracht at el. in PVAc
\cite{Spiess}. Concurrently with the publication of the theory or a
bit later, other types of non-linear studies have been performed on
several supercooled liquids which all quantitatively confirm the RFOT
prediction. These studies include experiments using nanometer
dielectric probes by Russell and Israeloff
\cite{RusselIsraeloff}. Deviations of the hydrodynamics of small
probes from the Stokes-Einstein relation \cite{Ediger_hydro} are also
consistent with this length scale as earlier predicted by Xia and
Wolynes \cite{XWhydro}, who showed deviations from the Stokes-Einstein
relation are expected for probes smaller the length $\xi$. These
deviations are especially notable in that they explicitly show the
cooperativity length is temperature dependent \cite{XWhydro}. Very
recently, evidence of a dynamic cooperative length, which increases
with lowering the temperature, have been obtained using a rigorous
inequality based on the non-linear susceptibility. Berthier at
el. \cite{Berthier}, again, decisively confirm the result in
Eq.(\ref{xi_univ}). Finally, while the detailed temperature dependence
of the cooperativity length is somewhat modified by the mentioned
barrier softening effects, the universality from Eq.(\ref{xi_univ}) is
robust, in the whole range of liquid fragilities \cite{LW_soft}.

Eq.(\ref{F(N)eq}) only gives the typical droplet nucleation profile.
Specific activation paths will reflect local variations of the liquid
landscape, i.e. they depend on the initial local configuration. These
variations are encoded in the local density of states (DOS) that can
be connected to the original configuration. A higher than average
density of states implies more configurations are available, leading
to a lower activation free energy barrier and a smaller number of
molecules participating in a structural transition, see
Eqs.(\ref{Fsc}) and (\ref{rsc}). Local variations in the DOS
correspond to fluctuations of the configurational entropy. Entropy
fluctuations, on the other hand, follow from the standard formula $\la
(\Delta S_c)^2 \ra = \Delta C_P$. Only the configurational part of the
heat capacity, as given by the heat capacity jump at the glass
transition, enters. Just this logic was followed in
Ref.\cite{XWbeta}. At this level of approximation, the entropy
fluctuations, which are approximately gaussian (surely, for $N^* >
6$), lead to a nearly gaussian distribution of barriers, via
Eq.(\ref{Fsc}). The ratio of the barrier distribution width to its
most probable value depends only on the liquid's fragility
\cite{XWbeta}:
\begin{equation} \label{dFF} \frac{\delta F^\ddagger}{F^\ddagger_\smp}
  \simeq \frac{1}{2\sqrt{D}}.
\end{equation}
As a result of the barrier distribution, {\em bulk} measurement of
relaxations will produce non-exponential time decay, or,
non-Lorentzian profiles in the frequency domain. The former is often
fitted with a stretched exponential: 
\begin{equation} \label{stretched}
  p(t) =  e^{-(t/t_0)^\beta}. 
\end{equation}
Assuming the barrier distribution is purely gaussian, one indeed
recovers a decay profile closely resembling the stretched exponential
form, where the corresponding exponent $\beta$ is related to the
fragility in the following simple manner \cite{XWbeta}:
\begin{equation}
  \beta \simeq \left[ 1 + (\delta F^\ddagger/k_B T)^2 \right]^{-1/2}.
\end{equation}

\begin{figure}[t]
\includegraphics[width=.95\figurewidth]{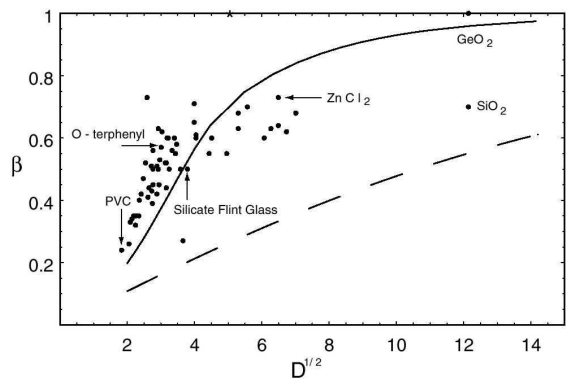}
\caption{\label{beta_D} This figure shows the correlation between the
  liquid's fragility and the exponent $\beta$ of the stretched
  exponential relaxations, as predicted by the RFOT theory,
  superimposed on the measured values in many liquids taken from the
  compilation in Ref.  \cite{Bohmer}. The dashed line is based on a
  simple gaussian barrier distribution, with the width mentioned in
  the text. The solid line takes into account the averaging effect of
  environmental rearrangments surrounding a mosaic cell, so that the
  barrier distribution to the right of the most probable value is
  replaced by a narrow peak of the same area; the peak is located at
  that most probable value.  This figure is taken from \cite{XWbeta}.}
\end{figure}

This correlation is shown by the dashed curve in
Fig.\ref{beta_D}. While already qualitatively consistent with
experimental data, further improvement in agreement is immediately
achieved upon realizing that if a particular region happens to be
relatively short on alternative structural states, a slightly
distinct, and hence overlapping (!), region is not likely to be so
disabled.  Since the environment of the original ``slow'' domain will
change such a slow region will relax sooner than expected. Xia and
Wolynes suggested therefore the barrier distribution, for
$\alpha$-relaxation, should be asymmetric, with relatively less weight
on the high barrier flank. The simplest way to incorporate the above
considerations into the theory is to use the original gaussian
distribution, but with the slow half of that distribution replaced by
a delta function carrying the same total weight, and centered at the
most probable value of the original gaussian. This approximation
introduces no new parameters. Using this simple form still leads to a
correlation of $\beta$ with $D$ and reproduces the empirically known
correlation between $\beta$ and $D$ quantitatively \cite{XWbeta}, see
the solid line in Fig.\ref{beta_D}. The RFOT theory clearly predicts
the degree of non-exponentiality is temperature dependent, consistent
with findings of Dixon and Nagel \cite{DixonNagel}.

\subsection{Dynamics near the Crossover}
\label{softening}

The temperature $T_A$ is a mean-field spinodal where one expects a
siginificant barrier softening, as $T_A$ is approached from
below. Above $T_A$ transport is no longer strictly activated. Instead, weakly correlated
collisions become the dominant contributor to the liquid's viscous
response. The barrier softening effects have been quantitatively
assessed by us \cite{LW_soft}. These effects are particularly important at
viscosities below 10 Poise or so, for all considered substances.

\begin{figure}[t]
  \includegraphics[width=.95\figurewidth]{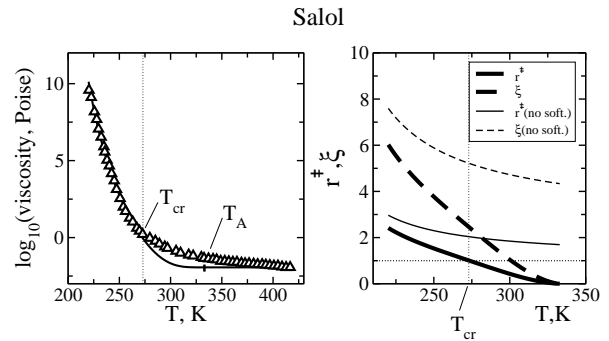} \vspace{2mm}
  \caption{\label{Salol1} Experimental data (symbols) for salol's
    viscosity \cite{Stickel}, superimposed on the results of the
    fitting procedure (line) from \cite{LW_soft} are shown. $T_A$ is
    the temperature at which the meanfield barrier vanishes, indicated
 by a tickmark. The configurational entropy, used in the
    fitting procedure, was extracted from experiment, see
    Ref.\cite{RichertAngell} and Eq.(\ref{scT}).  The temperature
    $T_\scr$ signifies a cross-over from activated to collisional
    viscosity, dominant at the lower and higher temperatures
    respectively.  The temperature is varied between the boiling point
    and the glass transition.  The r.h.s.  pane depicts the
    temperature dependence of the length scales of cooperative motions
    in the liquid. The thick solid and dashed lines are $r^\ddagger$
    and $\xi$ respectively. This figure is taken from \cite{LW_soft}.}
\end{figure}

For salol, the theoretically derived contribution to the relaxation
stemming from activated processes is shown as the solid line in the
left pane of Fig.\ref{Salol1}. The theory uses experimentally derived
values of the configurational entropy and the high temperature limit
of the relaxation times using the prefactor given by collision
theory. The softening correction computed by Lubchenko and Wolynes
(LW) relies on the idea that at small droplet radii, there is less
interface wetting. The correction terms in their analysis involve only
one fitting parameter, which is the spinodal dynamical crossover
temperature $T_A$, where the mean-field barrier between liquid states
would vanish. Clearly, at viscosities less than 10 Poise or so, where
the theoretical and the experimental curve bifurcate, the transport
becomes largely collisional. For several systems, the temperature
where the experiment and theory diverge, turns out to lie rather close
to the temperature $T_c$, where Stickel at el. \cite{Stickel} have
found a ``kink'' in the temperature dependences of relaxation
times. Often, two distinct VF forms have been used to fit the data in
the two temperature ranges, separated by $T_c$ \cite{Stickel}.

\begin{figure}[t]
  \includegraphics[width=0.8\figurewidth]{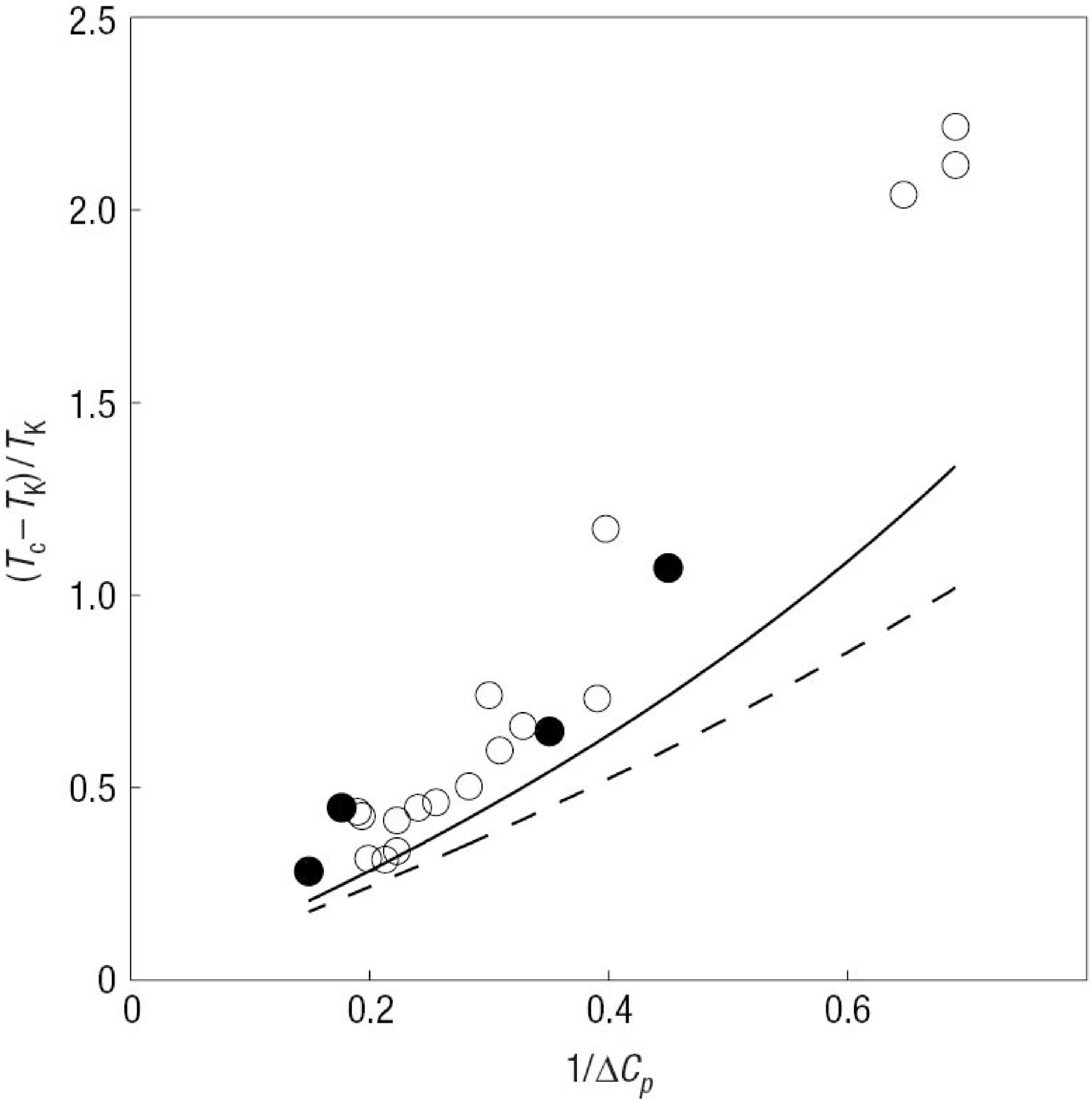}
  \caption{\label{SSW_Tc} From Ref.\cite{SSW}: Predictions for
    $(T_c-T_K)/T_K$ based on string-like rearranging regions is a
    dashed line, and $(T_c-T_K)/T_K$ based on percolation clusters is
    the solid line. The experimentally derived crossover temperatures,
    $(T_c^{exp}-T_K)/T_K$, from those materials collected by Novikov
    and Sokolov \cite{NovikovSokolovPRE}, are shown as circles with
    the dark circles referring to polymers.}
\end{figure}

While the LW analysis takes the temperature $T_A$ from fits, a recent
argument of Stevenson, Schmalian, and Wolynes \cite{SSW} provides a
microscopic description of the onset of activation-less
reconfigurational motions. The overall high free energy profile,
Eqs.(\ref{F(N)}) and (\ref{F(N)eq}), comes from the relative scarcity
of phase space trajectories that are sequences of low cost local
moves. The overall free energy cost should not only include
competition between the mismatch energy of mobilizing particles
against immovable neighbors and the entropy gain so achieved, but also
the number of different ways a connected pattern of mobile particles
can be placed on an aperiodic lattice. Keeping track of such patterns
can be done, for instance, by counting contiguous percolation clusters
\cite{Leath} or by counting impenetratable strings emanating from a
common origin. Here, again, the probability to find a contiguous
cluster grows with increasing configurational entropy. The estimates
based on percolation clusters or strings give distinct but comparable
values of the critical configurational entropy at which non-activated
reconfiguring short-circuits become possible \cite{SSW}: $s_c^{\tperc}
=1.28 k_B$ and $s_c^\tstring = 1.13 k_B$.  The corresponding
temperature at which a crossover to non-activated transport occurs, is
given by
\begin{equation}
  \frac{T_c^{\sperc}}{T_K} = \left(1 - \frac{s_c^\tperc}{\Delta c_p} 
    \frac{T_K}{T_g} \right)^{-1},
\end{equation}
A similar result is obtained for the ``string'' transition. Here, the
functional form for the configurational entropy from Eq.(\ref{scT})
has been used.  In Fig.\ref{SSW_Tc}, we show the theoretically derived
values of $T_c^\sperc$ and $T_c^\sstring$, in comparison with the
Stickel's $T_c$. Note that no adjustable parameters are used in
Fig. \ref{SSW_Tc}. It is interesting that the mathematics of the
string calculation is isomorphic to the Hagedorn transition in the
string theories of particle physics \cite{Hagedorn}.

We see that RFOT theory predicts deviations from the strict
Vogel-Fulcher dependence of the relaxation times on the
temperature. These deviations are expected to occur also at the faster
side of the full dynamical range of liquid relaxations. The deviations
span about three orders of magnitude, compared with the overall range
of $10^{17}$ or so, accessible in experiment. At these faster times,
according to Ref.\cite{SSW}, reconfigurations involving non-compact
regions become important. This is consistent with ``string-like''
excitations that had been observed in numerical simulations of liquid
dynamics by Glotzer and coworkers \cite{Glotzer_strings}.

Generally simulations are carried out in the dynamic range where these
softening effects are important. Yet, as in the last example, the RFOT
theory is indeed consistent with current simulational studies. To be
specific, current computational technologies are limited to a dynamic
range of nine decades or so. The time step in simulation is roughly a
femtosecond, significantly less than a typical collision time. In view
of the large system size necessary to avoid boundary effects, at best
one produces currently a microsecond run, implying only thermally
representative relaxations on a small fraction of a microsecond
time-scale can be ascertained.  (Equilibrating a liquid in a realistic
simulation, in the deeply supercooled regime, must of course be
subject to the very same high barriers present in the laboratory
melts.)  According to a recent discussion of Lubchenko \cite{L_JNC},
such short times imply the size of the critical nucleus (or the
participation number at the saddle point, if you will) is only about
$(4 \pi/3) 1^3 \simeq 5$ ``beads'', corresponding to $r^\ddagger/a
\simeq 1$.  In silica, the bead is a fraction of a SiO$_4$ tetrahedron
\cite{LW_soft, LW_RMP}, implying a critical region size of 15 atoms or
so. This is consistent with the participation ratios, at the barrier
top, reported recently by Reinisch and Heuer in simulations
characteristic of these short times \cite{ReinischHeuer}. According to
this discussion, one must be careful so as not to mistake a possible
but unrepresentative high barrier trajectory, that could be always
found given a high enough temperature, for a thermally representative
trajectory that actually describes reconfiguration on the laboratory
time scale.

\subsection{Aging}

\begin{figure}[t]
 \includegraphics[width=.6\figurewidth]{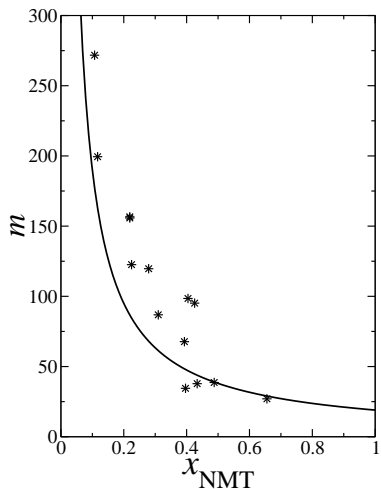}
 \caption{\label{m_x} The fragility parameter $m$ is plotted as a
   function of the NMT nonlinearity parameter $x_{\sNMT}$. The curve
   is predicted by the RFOT theory when the temperature variation of
   $\gamma_0$ is neglected. The data are taken from
   Ref. \cite{app_phys_rev}.  The more fragile substances consistently
   lie above the prediction, which has no adjustable parameters. This
   discrepancy may be due to softening effects.}
\end{figure}

A liquid is in equilibrium. Ultimately there is no memory of its
initially prepared configuration. But when the liquid is cooled faster
than it can equilibrate, the system will find itself in a {\em subset}
of all the states available before the quench began. Which subset
depends on the thermal history. The resultant quenched glass is a
truly non-ergodic system. It will begin to relax towards the
thermodynamic state which would have been typical upon slower
quenching. But unlike at temperatures above vitrification, the
structural re-arrangement below $T_g$ no longer depends only on the
ambient temperature $T$ alone but also on the temperature history. The
vibrational temperature is near ambient, while the (now mostly static)
structure is representative of an equilibrated sample at a ``fictive''
temperature $T_f$. Strictly, one must detail the complete temperature
schedule. Yet, since the relaxations near a glass transition on the
routine laboratory scales are so slow - seconds and slower - that most
quenches would result in a structure typical of a single temperature
$T_g$, save the smaller vibrational amplitudes. This immediately
implies the energy barrier distribution should be nearly temperature
independent.  If one further replaces the full barrier distribution by
a single typical barrier, one expects the temperature dendence of
relaxation times in frozen glasses to follow a simple Arrhenius law. A
similar expectation serves as the basis of the phenomenological
framework of Nayaranaswany-Moynihan-Tool:
\begin{equation}
k_{n.e.} = k_0 \exp\left\{-x_{NMT}\frac{\Delta E^*}{k_B T} -
(1-x_{\sNMT}) \frac{\Delta E^*}{k_B T_f} \right\}. \hspace{4mm}
%\label{k_AG}
\end{equation}
where $E^*$ is the equilibrated apparent activation energy at $T_g$
and $x_{\sNMT}$ lies between 0 and 1.  The general expression for
activated transitions within RFOT theory in Eq.(\ref{F(N)}), which
allows one to treat transitions from an arbitrary initial local state
and implies the apparent activation energy below $T_g$ is
significantly smaller than above the glass transition. The free energy
difference $f_{in} - f_{eq}$ becomes nearly temperature independent
below $T_g$, also consistent with the initial state energy being {\em
  above} the equilibrated value. On the other hand, the rate of the
temperature change of the apparent activation energy above $T_g$
depends on the temperature dependence of the configurational entropy,
see Eq.(\ref{Fsc}). The RFOT theory thus predicts the nonlinearity
parameter $x_{\sNMT}$ to be correlated with the fragility. A
straightforward calculation yields the following simple relation
\cite{LW_aging}:
\begin{equation}
m \simeq \frac{19}{x}.
\end{equation}
This prediction is compared to experimental data in Figure \ref{m_x}. (Here, we
set $T_f = T_g$.)
In addition to predicting the typical activation energy in frozen
glasses, the RFOT theory can be employed to estimate the spread in the
barrier distribution, and thus enable one to compute the temperature
dependence of the exponent $\beta$ of the stretched exponential
relaxation, as in Eq.(\ref{stretched}). Corresponding experimental
data addressing this prediction so far appear inconclusive
\cite{Alegria, LehenyNagel}.

\section{Quantum Theory of Glasses}
\label{Cryo}

The microscopic picture of structural relaxations in supercooled
liquids and glasses built up from the RFOT theory provides a starting
point for understanding a radically different dynamical regime,
i.e. the behavior of structural glasses at cryogenic temperatures \cite{LW},
\cite{LW_BP}, \cite{LW_RMP}, \cite{LSWdipole}. As we have recently
provided a detailed account of these findings elsewhere \cite{LW_RMP,
  LSWdipole}, we only summarize here several important points which
help to strengthen the case for the RFOT theory predictions near
$T_g$.

Much below the Debye temperature $T_D$, cryogenic glasses exhibit a
set of excitations in excess of the usual Debye density of states.  At
temperatures below $1$K or so, these lead to a linear heat capacity
and enhanced phonon scattering. In a higher temperature range around
10 K or so ($\sim 1$ THz), yet another pronounced feature in the
apparent density of excitations, called the ``Boson Peak'', is
observed \cite{LowTProp}. This excess density of states also rapidly
shortens the mean free path of thermal phonons.

The mosaic predicted by RFOT theory immediately implies there are
local structural excitations that may be viewed as local multi-level
systems. These anharmonic motions would interact with the
mostly elastic lattice.  At low enough temperatures, one expects only
the two lowest energy levels to contribute thermodynamically (but see
below). This is in line with the early, phenomenological two-level
system (TLS) model which empirically describes the low T regime
\cite{AHV, Phillips}. But the low energy tail of the structural
excitations can be computed from first principles. First one finds it
is nearly energy independent at the lowest energies, as in the
phenomenology. The corresponding density of states $\bar{P}$ is given
by a simple expression \cite{LW}:
\begin{equation} \label{Pbar} \bar{P} \simeq \frac{1}{k_B T_g \xi^3},
\end{equation}
This gives a linear heat capacity whose magnitude is consistent with
experiment. The condition of marginal stability of structural
excitations to lattice distortions, at the glass transition, allows
one to derive the coupling of those frozen-in transitions to the
phonons:
\begin{equation} \label{g} g \simeq \sqrt{k_B T_g \rho c_s^2 a^3},
\end{equation}
where $\rho$ and $c_s^2$ are the mass density and the speed of sound
of the glass.  These two expressions, representing the nature of the
states frozen at $T_g$, enable us to understand the mysterious
universality of phonon scattering, found in many amorphous insulators:
The ratio of the phonon mean free path to the phonon wave length
depends on $\bar{P} g^2$ and turns out to be about 150 for all
substances. Thus the standard tunneling model implies $\bar{P}
g^2/\rho c_s^2 \sim 10^{-2}$ \cite{FreemanAnderson}. Combining
Eqs.(\ref{Pbar}) and (\ref{g}) from RFOT theory shows that the
universality is expected because it reflects the nearly universal
value of the cooperativity length at the glass transition \cite{LW}:
\begin{equation}
  \bar{P} g^2/\rho c_s^2 \simeq \left(\frac{a}{\xi}
  \right)^3_{T_g} \simeq1/200,
\end{equation}
Another simple correlation empirically found but otherwise unexpected
follows from RFOT theory. It relates the TLS-phonon coupling $g$ to
the glass transition temperature $T_g$ via the Lindemann ratio:
\begin{equation}
  g \simeq \frac{T_g}{(d_L/a)} \simeq 10. k_B T_g.
\end{equation}
At the very lowest temperatures, the structural transitions involve
the {\em tunneling} between two states of domain wall traversing a
local region, spanning a length $\xi$ across. But the RFOT theory
suggests these motions can support more than two states. At
temperatures above 1 K or so, higher energy states of these tunneling
centers are predicted to come into play. Microscopically, these
motions can be visualized as vibrational excitations of the domain
walls. The spectrum of these ``capillary waves'', or ``ripplons'', can
be computed without adjustable parameters. The corresponding
frequencies and density of states quantitatively account for the
apparent excess heat capacity and phonon scattering in the terahertz
range, that have been associated with the Boson Peak.  The simplest
quantitative estimate of the vibration frequencies is obtained
neglecting interaction with the phonons and yields for the Boson Peak
frequency \cite{LW_BP}:
\begin{equation}
  \omega_\sBP \simeq (a/\xi) \omega_D,
\end{equation}
where $\omega_D$ is the Debye frequency. Effects of ripplon-phonon
interaction will lead to a shift and brodening of the ripplon
resonances, and seem to explain the non-universality of the thermal
conductivity plateau.  The multilevel nature of ``two level systems''
in this temperature range is also apparent in single molecule
spectroscopy experiments when complex spectral trails have ben
observed by Orrit and coworkers \cite{Orrit}.

The tunneling centers (TC) are fluctuating resonances and hence will
mutually attract, following the usual London dispersion law
$r^{-6}$. Since the number of active centers grows with temperature,
this implies the inter-center attraction provides a mechanism for an
additional attracting holding the solid together at higher
temperatures. Like rubber, this yields a negative thermal expansion
coefficient!  Quantitative estimates \cite{LW_RMP} show that the
attraction is greatly enhanced by the presence of ripplons, even at
very low $T$. This mechanism yields a magnitude of the Gr\"{u}neisen
parameter consistent with experiment, in a number of glasses, which is
often very anomalously large and negative \cite{Ackerman}.

Finally, individual molecular motions during the structural
transitions will result in local electric charge
redistribution. Ultimately this leads to a coupling of the transitions
with electromagnetic waves. Recent estimates \cite{LSWdipole} show the
transition induced electric dipole is about a Debye in magnitude, in
spite of there being several hundred atoms participating in the
motion. This value is consistent with experiments \cite{Maier,
  Kharlamov_dipole}.

\section{Concluding Remarks - Future Directions}
\label{Summary}

In place of a summary, we discuss the range of applicability of the
RFOT theory; the limitations uncovered naturally suggest directions
for future efforts.

We have already pointed out that the RFOT theory's predictions of the
relaxation time and size scales are quantitative at viscosities of 10
Poise and above. This range of time scales constitutes, on the
logarithmic scale, at least 80\% of the routinely probed dynamic range
in liquids of $10^{-2}$ to $10^{14}$ Poise or so. The ability of the
theory to make predictions here relies on the fact that the motions
are mesoscopic, hundreds of atoms being reconfigured, allowing
sufficient thermodynamic averaging.

Based on mesoscale averaging the theory thus applies only when there
is no significant medium range order. Many liquids seem to satisfy
this constraint, and span a wide range fragilities and chemical
properties. These include nearly ionic compounds, such SiO$_2$,
ZnCl$_2$; alcohols, such as propanol; aromatic and aliphatic
hydrocarbons such as trinaphthylbenzene or
2-methylpentane. Measurements on such simple liquids confirm the
predictions of the RFOT theory. The list of substances quantitatively
described presently includes several dozens of materials and keeps
growing.

On the other hand, some substances, while showing most of the
signatures of glassy behaviors, exhibit additional types of order and
scales of motion that will require separate treatment. We have in fact
mentioned several examples of such systems: They include many of the
most important polymers. In polymers, for instance, it is clear that
various chain motions may provide long length scales and long
time-scale relaxations, especially when the persistence length is
large. Even for such systems, one still finds qualitative agreement
with the RFOT calculations. For example, most polymers show TLS and
Boson Peak densities of states that are comparable to those of simple
molecular systems. One consequence of the possiblity of mid-range
order is that one should exercise caution in searching for
universalities in the glassy behaviours. The RFOT theory, among other
things, provides in fact a first principles basis for such
searches. As we have seen, substances that conform to the RFOT
specifications, are predicted to show not one but rather a whole set
of correlations, most of which we have mentioned earlier: $T_K$
vs. $T_0$, $\Delta c_p$ vs.  $D$ (or $m$), $\xi$ vs. $T$, $\xi(T_g)$
vs. $a$, universality of $s_c(T_g)$, $\beta$ vs. $D$, $\beta$ vs. $T$,
$m$ vs. $x$, $T_c$ vs. $\Delta c_p$, $\bar{P}$ vs. $T_g$, $\bar{P}$
vs. $\xi$ (or $a$), $\bar{P}$ vs. $g$, $\omega_{\sBP}$ vs. $\omega_D$,
$\omega_\sBP$ vs. $a$. The experiments become more difficult as one
proceeds down the list, so fewer comparisons of predictions with
experiment are available for the last few relations, but in our view
the number of confirmed relationships already provides a secure basis
to suggest anomalies are a sign of new degrees of freedom.

We were careful to limit the title of this review calling it ``Theory
of Structural Glasses and Supercooled Liquids'', not dealing with all
glassy systems. The quantitative applicability of RFOT theory to
molecular glassy systems can be ultimately traced back to an emergent
small parameter, the Lindemann ratio. Also for structural glasses, the
glass transition temperature $T_g$ is much below the dynamic
transition at $T_A$. This was pointed out clearly in an analysis by
Eastwood and Wolynes \cite{EastwoodW} who developed a ``Ginzburg''
criterion for the theory. Their criterion shows many exciting systems
exhibiting signs of glassiness are not in the appropriate universal
regimes for strict quantitation because those systems are
intrinsically softer with larger effective local motions. Colloidal
glasses, owing to the larger size of their constituents, are
intrinsically slow at the one particle level and thus at human
measurement time scales, are always near $T_A$. Gels, stripe glasses
\cite{SchmalianWolynes2000}, and microemulsions \cite{SWmayo} have
larger Lindemann ratios and although they are described by the RFOT
theory at the mean-field level, again, we expect there will be very
significant renormalization of their behavior in the activated regime.
Finally, although the RFOT theory provides a route from input
intermolecular forces to the macroscopic behavior, the present
quantitative successes of the theory avoid the hardest part of the
detailed microscopic modeling by utilizing the fact that it is
relatively easy to estimate configurational entropy from laboratory
measurements. As usual, completely ab initio calculations will be more
difficult, justifying greater attention to microscopic liquid state
theory and molecular simulation technology.

{\em Acknowledgments:} V.L. has been funded in part by the GEAR
Program and the New Faculty Grant at University of Houston.  The work
of P.G.W. is supported by the NSF grant CHE 0317017.

%\bibliography{/Users/vas/Documents/tex/ACP/lowT}

%\bibliographystyle{ieeetr} 
\bibliography{/Users/vas/Documents/tex/ACP/lowT}

\end{document}